\documentclass[%
 reprint,
superscriptaddress,
 amsmath,amssymb,
 aps,
 prb,
]{revtex4-2}
\usepackage{subfigure}
\usepackage{graphicx}
\usepackage{dcolumn}
\usepackage{bm}
\usepackage{booktabs}
\usepackage{amsmath}

\begin{document}


\title{Nonlinear Hall Effect in KTaO$_3$ Two-Dimensional Electron Gases}

\author{Patrick W. Krantz}
\affiliation{Department of Physics, Northwestern University, Evanston, Illinois. 60208, USA
}%
\author{Alexander Tyner}%
\affiliation{Graduate Program for Applied Physics, Northwestern University, Evanston, Illinois. 60208, USA
}%
\author{Pallab Goswami}
\affiliation{Department of Physics, Northwestern University, Evanston, Illinois. 60208, USA
}%
\affiliation{Graduate Program for Applied Physics, Northwestern University, Evanston, Illinois. 60208, USA
}%
\author{Venkat Chandrasekhar}
\affiliation{Department of Physics, Northwestern University, Evanston, Illinois. 60208, USA
}%
\affiliation{Graduate Program for Applied Physics, Northwestern University, Evanston, Illinois. 60208, USA
}%

\date{\today}

\begin{abstract}
The observation of a Hall effect, a finite transverse voltage induced by a longitudinal current, usually requires the breaking of time-reversal symmetry, for example through the application of an external magnetic field or the presence of long range magnetic order in a sample.  Recently it was suggested that under certain symmetry conditions, the presence of finite Berry curvatures in the band structure of a system with time-reversal symmetry but without inversion symmetry can give rise to a nonlinear Hall effect in the presence of a probe current.  In order to observe the nonlinear Hall effect, one requires a finite component of a so-called Berry dipole along the direction of the probe current.  We report here measurements of the nonlinear Hall effect in two-dimensional electron gases fabricated on the surface of KTaO$_3$ with different surface crystal orientations as a function of the probe current, a transverse electric field and back gate voltage.  For all three crystal orientations, the transverse electric field modifies the nonlinear Hall effect.  We discuss our results in the context of the current understanding of the nonlinear Hall effect as well as potential experimental artifacts that may give rise to the same effects. 
\end{abstract}

\maketitle

\section{Introduction}

The Hall effect is one of the best-known phenomena in condensed matter physics, having been discovered by Edwin Hall more than 140 years ago.  It manifests itself in simple conductors as a transverse voltage in response to a driving current and a perpendicular magnetic field.  It can also be observed in magnetic materials in the absence of an external magnetic field, the so-called anomalous Hall effect \cite{nagaosa_anomalous_2010}.  Since its discovery, the Hall effect has been studied intensely both experimentally and theoretically in a variety of materials \cite{karsenty_comprehensive_2020}, with quantized versions of both the ordinary Hall effect\cite{von_klitzing_40_2020} and the anomalous Hall effect \cite{he_topological_2018} now well established.  A unified picture of these various Hall effects can now be understood based on the underlying topology of the band structure of a material in the presence of external fields and currents \cite{senthil_symmetry-protected_2015}.  In this picture, the nature of the Berry curvature $\boldsymbol{\Omega}_n=\nabla \times \boldsymbol{A}_n$ of the bands is particularly important.  Here $\boldsymbol{A}_n (\boldsymbol{k}) = i <u_{n\boldsymbol{k}}|\nabla_{\boldsymbol{k}}|u_{n\boldsymbol{k}}>$ is the so-called Berry connection \cite{xiao_berry_2010}, with $u_{n\boldsymbol{k}}$ being the Bloch wavefunction of the band with index $n$ and crystal momentum $\boldsymbol{k}$. 

The Hall response is related to the integral of the Berry curvature $\boldsymbol{\Omega}(\boldsymbol{k})$ over the entire Brillouin zone. In a system with time reversal symmetry, one can show that the Berry curvature is an odd function of $\boldsymbol{k}$, $\boldsymbol\Omega(\boldsymbol{k}) = -\boldsymbol\Omega(-\boldsymbol{k})$, while in a system with space inversion symmetry, the Berry curvature is an even function of $\boldsymbol{k}$, $\boldsymbol\Omega(\boldsymbol{k}) = \boldsymbol\Omega(-\boldsymbol{k})$.\cite{xiao_berry_2010}  Consequently, in a system with both time reversal and space inversion symmetry, $\boldsymbol{\Omega}(\boldsymbol{k})$ identically vanishes over the entire Brillouin zone, and there is no Hall response.  In a system with time reversal symmetry, but broken space inversion symmetry, the Berry curvature is odd in $\boldsymbol{k}$ so that its integral over the Brillouin zone again vanishes and there is no resulting Hall voltage. Thus, broken time-reversal symmetry appears to required be to observe a finite Hall voltage.

The situation is different if there is an electric field $\boldsymbol{E}$ applied to the sample, as might be generated by a measurement current. In this case one obtains a Hall response that is quadratic in $E$, i.e., a nonlinear Hall effect.\cite{sodemann_quantum_2015}  The velocity of an electron in a band with index $n$ is given by\cite{sundaram_wave-packet_1999}
\begin{equation}
v_n(\boldsymbol{k}) = \frac{1}{\hbar} \nabla_{\boldsymbol{k}} \mathcal{E}_n - \frac{d \boldsymbol{k}}{dt} \times \boldsymbol{\Omega}_n (\boldsymbol{k})
\label{eqn:eqn1}
\end{equation} 
where the first term is the usual Bloch velocity of a band electron, and the second term is the anomalous velocity due to the Berry curvature.  In a 2D crystal, $\boldsymbol{\Omega}$ points out of the 2D plane, so that the anomalous velocity is in the plane of the crystal in a direction perpendicular to $d\boldsymbol{k}/dt$.  In the presence of an electric field $\boldsymbol{E}$, the time evolution of $\boldsymbol{k}$ is given by $d\boldsymbol{k}/dt = -e \boldsymbol{E}$ in the semiclassical approximation.  The current arising from the anomalous velocity is given by \cite{moore_confinement-induced_2010}
\begin{equation}
\boldsymbol{j} = \frac{e^2}{\hbar} \int \frac{d^2k}{4 \pi^2}  \bigl(\boldsymbol{E} \times \boldsymbol{\Omega}(\boldsymbol{k})\bigr) g(\boldsymbol{k})
\label{eqn:eqn2}
\end{equation}
where $g(\boldsymbol{k}) = f(\boldsymbol{k}) - f_0(\boldsymbol{k})$ is the deviation of the electron distribution function $f(\boldsymbol{k})$ from its equilibrium value $f_0(\boldsymbol{k})$ due to perturbations from the driving field, $\boldsymbol{E}$.  To first order in the low frequency limit, $g(\boldsymbol{k})$ is given by \cite{ziman_1972}
\begin{equation}
g(\boldsymbol{k}) = \left(- \frac{\partial f_0}{\partial \mathcal{E}}\right) \tau v_{\boldsymbol{k}} \cdot e \boldsymbol{E}
\end{equation}   
where $\tau$ is the scattering time and $v_{\boldsymbol{k}}$ is the first term in Eqn. (\ref{eqn:eqn1}).  The derivative of the electron distribution function ensures that the primary contribution to the current in Eqn. (\ref{eqn:eqn2}) comes from the Fermi surface.  We have also dropped the band index in the equations above, simplifying to the case where only one band is important.  One can see immediately that the current arising from the anomalous velocity is perpendicular (transverse) to $\boldsymbol{E}$, and that it is quadratic in $\boldsymbol{E}$, as noted earlier.

Recalling that $v_{\boldsymbol{k}} = (1/\hbar) \nabla_{\boldsymbol{k}} \mathcal{E}$, we can rewrite $-(\partial f_0/\partial \mathcal{E}) v_{\boldsymbol{k}}$ as  $-(1/\hbar)  \nabla_{\boldsymbol{k}} f_0$.  Taking all the $k$ independent terms of Eqn. (\ref{eqn:eqn2}) out from under the integral, and writing it in scalar form, we obtain an expression for the transverse current in response to the electric field $E$
\begin{equation}
\boldsymbol{j} = \frac{e^3 \tau}{\hbar^2} E^2 \int \frac{d^2k}{4 \pi^2 } (\nabla_{\boldsymbol{k}} f_0) \Omega_k.
\end{equation}   

Integrating by parts, one obtains
\begin{equation}
\boldsymbol{j} = \frac{e^3 \tau}{\hbar^2} E^2 \int \frac{d^2k}{4 \pi^2 } f_0 (\nabla_{\boldsymbol{k}} \Omega_k)
\label{QuantumHallCurrent}
\end{equation}
assuming that $\Omega_k$ is odd in $k$.

\begin{figure}
\includegraphics[width=8cm]{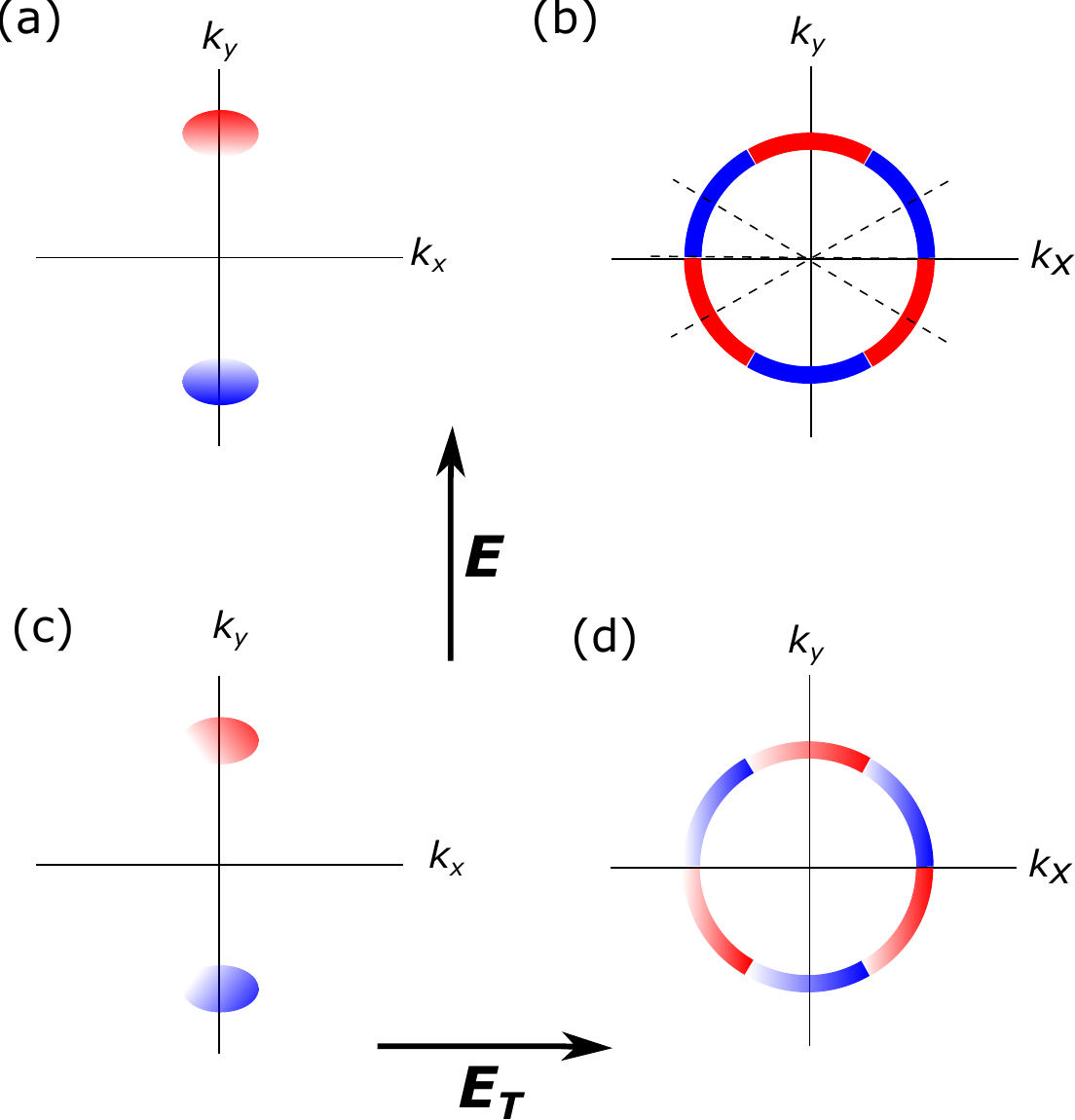}
\caption{Schematic of a Berry curvature in a 2D crystal.  Red denotes a positive Berry curvature, while blue denotes a negative Berry curvature.  (a)  A crystal with $C_2$ symmetry with a single mirror line along $k_y$, with the electric field $\boldsymbol{E}$ applied along $k_y$.  (b) A crystal with $C_6$ symmetry with three mirror lines, shown as the dotted lines (one is along the $k_y$ axis).  Modification of the Berry curvature of (a) with an additional electric field $\boldsymbol{E}_T$ applied along the $k_x$ direction, perpendicular to $\boldsymbol{E}$.  (d) Similar modification of the Berry curvature of (b) with $\boldsymbol{E}_T$.  The application of $\boldsymbol{E}_T$ destroys the mirror symmetries.} 
\label{fig:berrycurvatureschematic}
\end{figure}

Sodemann and Fu \cite{sodemann_quantum_2015} identify the integral in Eq. \ref{QuantumHallCurrent} as the dipole moment of the Berry curvature, or Berry dipole. The resulting current is proportional to the Berry dipole, and responds to experimental handles that modify the stength of the dipole. Sodemann and Fu also note that the symmetry considerations for such a Berry dipole to exist (and hence for a nonlinear Hall effect to be observable) in a two-dimensional (2D) crystal are rather stringent:  the crystal must have only a single mirror line, hosting the dipole perpendicular to that line \cite{sodemann_quantum_2015}.  Figure \ref{fig:berrycurvatureschematic} shows a schematic representation of a putative Berry curvature in two different 2D crystals.  The first one shown in Fig. \ref{fig:berrycurvatureschematic}(a) represents a crystal that has a single mirror line with a finite Berry curvature and a Berry dipole  aligned along the $k_y$ axis.  The schematic of Fig. \ref{fig:berrycurvatureschematic}(b) represents a crystal with a three-fold rotation symmetry and three mirror lines, similar to the (111) KTO surface. Taking the integral over the full Brillouin zone as described in Eq. \ref{QuantumHallCurrent} yields no net Berry dipole in any direction. Thus the crystal of Fig. \ref{fig:berrycurvatureschematic}(a) will show a nonlinear Hall effect if an external electric field $\boldsymbol{E}$ is applied in the direction shown, while the crystal of Fig. \ref{fig:berrycurvatureschematic}(b) will not.

  The application of an additional electrical field $\boldsymbol{E}_T$ will modify an existing Berry dipole, as shown in Fig. \ref{fig:berrycurvatureschematic}(c). In a system where there is no net Berry dipole due to the existence of multiple mirror symmetry lines, the application of $\boldsymbol{E}_T$ along one of these mirror symmetry lines can result in a net Berry dipole by effectively removing the remaining mirror symmetries, as shown schematically in Fig. \ref{fig:berrycurvatureschematic}(d) for the crystal of Fig. \ref{fig:berrycurvatureschematic}(b).  Consequently, one expects to observe a nonlinear Hall effect in such a system that depends on the magnitude of the transverse electric field $\boldsymbol{E}_T$.  Such a field dependent nonlinear Hall effect has recently been reported in WTe$_2$ \cite{zhang_high_mobility_2018}.  For devices in which a nonlinear Hall effect is observed in the absence of $\boldsymbol{E}_T$, the application of a transverse field might be expected to modify the magnitude of the observed effect, as shown schematically in Fig. \ref{fig:berrycurvatureschematic}(c).  Note that even if $\boldsymbol{E}$ and $\boldsymbol{E}_T$ are not along the crystal axes of the crystal, one would still expect to observe a nonlinear Hall effect, although its dependence on $\boldsymbol{E}_T$ might be more complicated.

The quadratic dependence of the transverse current on the driving electric field enables an elegant way to isolate the nonlinear Hall effect, as emphasized by Sodemann and Fu \cite{sodemann_quantum_2015}. If an ac measuring current $I_\omega$ is used at a (low) frequency $\omega$, then the resulting nonlinear Hall voltage $V_T$ will have components at dc and $2\omega$.  As thermoelectric and other effects may also result in dc contributions that are difficult to isolate, it is easier to focus on the $2\omega$ component of the Hall voltage, $V^{2\omega}$, which should have a quadratic dependence on the measuring current, $V^{2\omega}\propto I_\omega^2$ for low values of $I_\omega$.  However, it is important to note that there are other contributions that may result in a $2\omega$ signal that is quadratic in $I_\omega$.  In particular, if the response $V_T(I)$ is nonlinear in $I$, then a standard ac lock-in measurement will result in a $2\omega$ signal that is quadratic in $I_\omega$.  Many materials that are not necessarily topological might have nonlinear current-voltage characteristics.  Indeed, as observed by Webb, Washburn and Umbach \cite{webb_PhysRevB.37.8455} almost four decades ago, even devices made from conventional metals such as gold have significant response at harmonics of the probe current frequency.  Thus it is important to identify characteristics of the second harmonic response that relate the observed signal to Berry curvature effects.

We describe below the results of our measurements of the nonlinear Hall effect in KTaO$_3$ (KTO) two-dimensional electron gases (2DEGs) with three different crystal surface orientations: (001), (110) and (111).  We report measurements of both the first harmonic ($V^{\omega})$ and the second harmonic ($V^{2\omega})$ longitudinal and transverse voltage responses as a function of the ac drive current $I_\omega$, backgate voltage $V_g$ as well as an additional dc current $I_{dc}$ applied between the transverse voltage probes.  The (001) and (111) crystal orientations nominally have more than one mirror line, so that according to the discussion above, no nonlinear Hall effect should be observed in the absence of the transverse current.  For all three crystal orientations, however, we observe a second harmonic transverse voltage that is modified by $I_{dc}$.  This transverse second harmonic signal is smallest for the (001) oriented sample, larger for the (110) sample and largest for the (111) sample.  We discuss our results in terms of the current understanding of the nonlinear Hall effect and potential experimental artifacts.

\section{Sample fabrication and measurement}
\subsection{Sample fabrication}
The devices used in this work were Hall bars patterned using photolithography on 5 mm x 5 mm x 0.5 mm KTO single-side polished crystal substrates obtained from MSE Supplies LLC in three different surface crystal orientations: (001), (110) and (111).  Prior to coating with photoresist, the substrates were subjected to a standard cleaning regimen (deionized water, acetone, and isopropyl alcohol) before being annealed at 650 C for two hours in an ambient environment followed by two hours anneal in deionized water as described by Tomar \textit{et al.} \cite{tomar_realization_2018} for optimal TaO surface termination.  In an earlier study, this last annealing step was not performed, resulting in devices that did not go superconducting down to our lowest measurement temperatures.  For the present study, the (110) and (111) oriented devices did go superconducting at a maximum temperature of $\sim$1.1 K.  

In order to create the 2DEGs, the substrates were first patterned into 4 standard six-terminal Hall bar geometries, each of length 600 $\mu$m and width 50 $\mu$m and then metallized in a e-gun evaporator with 99.9995\% Al.  Prior to deposition, the samples were cleaned \textit{in situ} with 100 mT oxygen in order to remove any photoresist residue.  After the plasma cleaning, 1.5 nm of Al was deposited and the substrate was allowed to sit for 10 minutes under vacuum in order to getter oxygen from the substrate surface.  This was followed by a second deposition of 1.5 nm of Al, after which 100 mTorr of O$_2$ was introduced into the evaporator and the sample allowed to sit for another 10 minutes in this atmosphere.  A final layer of 2 nm of Al was then evaporated and oxidized.   Measurements of a co-evaporated glass slide confirmed that the deposited metal was not conducting.  All three crystal orientations were processed simultaneously.

Substrates supplied by commercial suppliers are usually cut along well defined crystal directions.  During the photolithography exposure, the Hall bar mask was aligned so that the lengths of the Hall bars were parallel to the edges of the substrate with an accuracy of better than 1$^\circ$.  For the (001) surface, this meant that the Hall bars were aligned along the equivalent $<100>$ surface directions.  For the (111) surface, two of the Hall bars were aligned along the $[1\bar{1}0]$ and two along the $[1\bar{1}2]$ direction: we show data below for a Hall bar aligned along the $[1\bar{1}0]$ direction. All of these alignments were verified by Laue diffraction, though Laue did reveal that the (110) oriented surface crystal substrate had \emph{not} been cut along the relevant cubic crystal directions ($[1\bar{1}0]$ and $[001]$), but at an angle of 45$^\circ$ to these axes, hence the Hall bars were aligned at this angle with respect to the principal crystal directions. 

\begin{figure}
	\includegraphics[width=8cm]{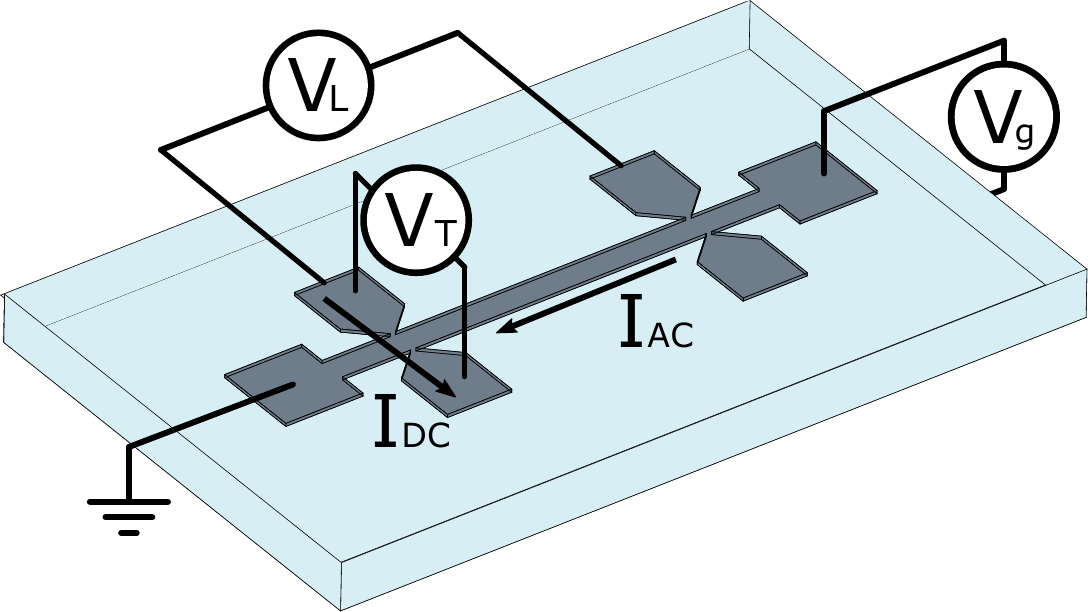}
	\caption{Schematic of the measurement configuration.} 
	\label{fig:sampleschematic}
\end{figure}

\begin{figure*}
	\includegraphics[width=18cm]{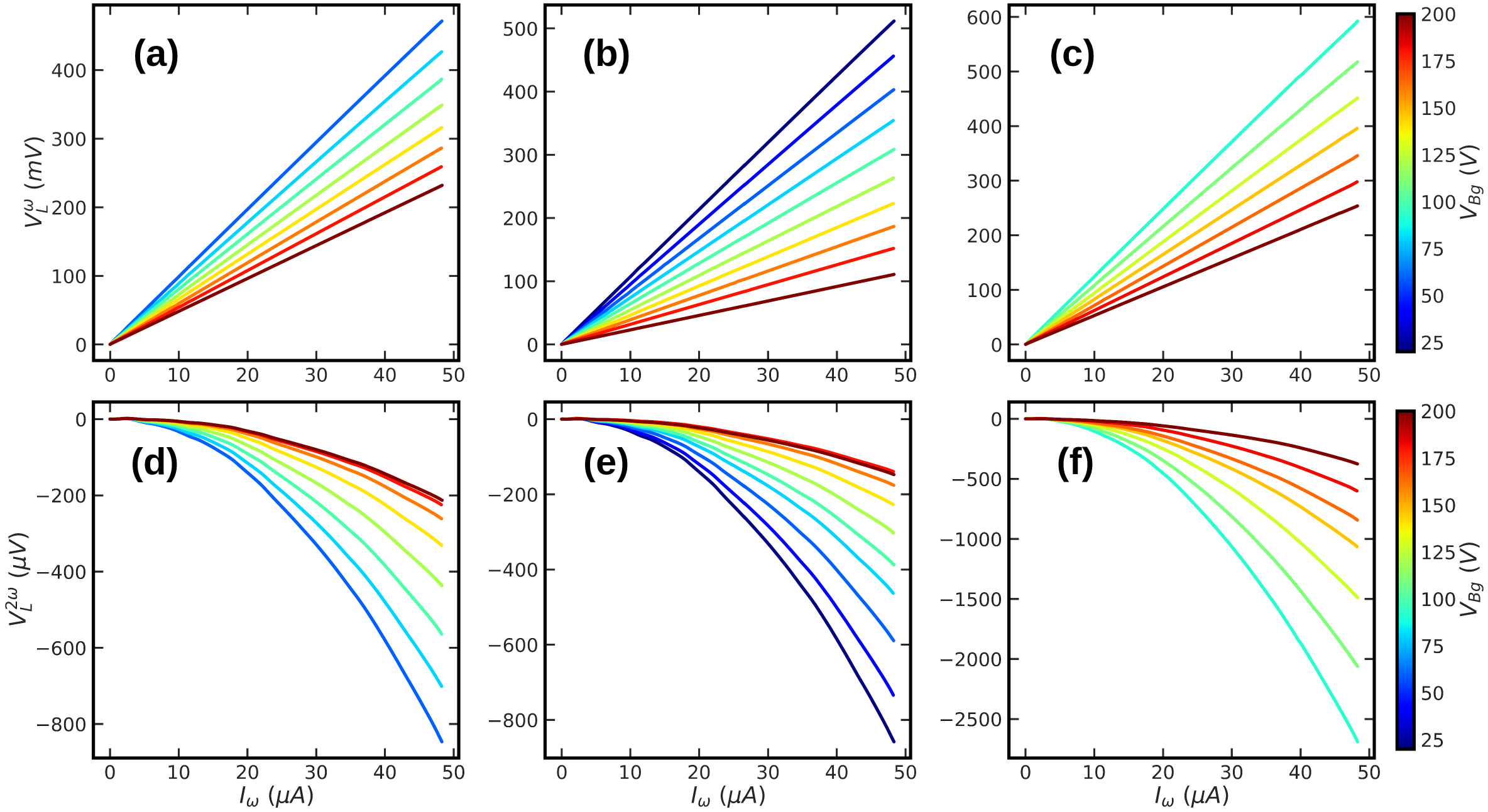}
	\caption{(a-c)  Longitudinal voltage response at the frequency $\omega$, $V_L^\omega$, as a function of ac drive current $I_\omega$ at a fixed transverse current $I_{dc}$ of 1 $\mu$A for various gate voltages $V_g$ for the (001) (a), (110) (b) and (111) (c) oriented devices. (d-f)  Simultaneously measured $V_L^{2\omega}$ vs. $I_\omega$.}      
	\label{fig:longitudinaldependenceA}
\end{figure*} 
\begin{figure*}
	\includegraphics[width=18cm]{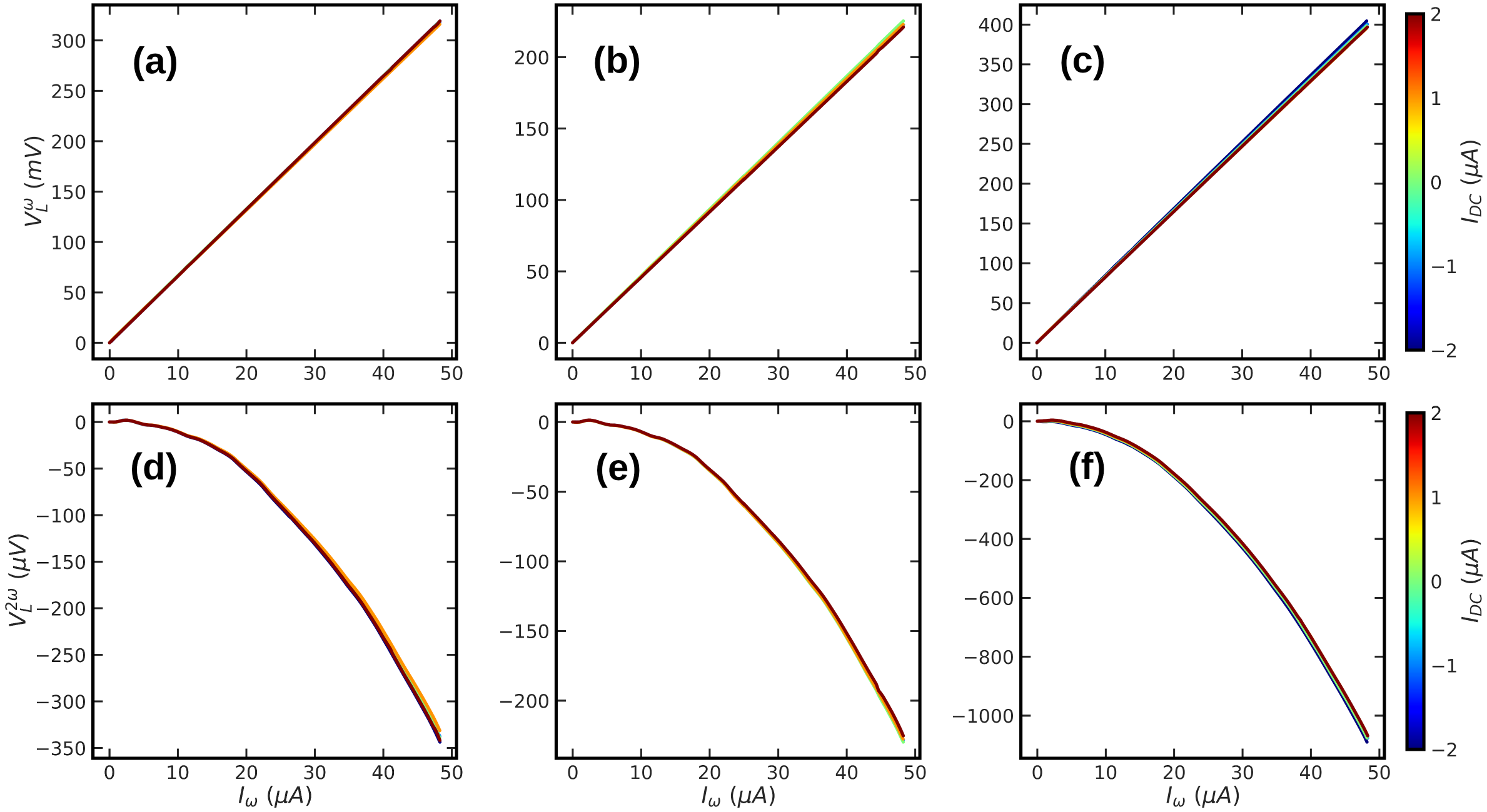}
	\caption{(a-c) Longitudinal voltage response at the frequency $\omega$, $V_L^\omega$, as a function of ac drive current $I_\omega$ at a fixed back gate voltage of $V_g$=140 V at various values of the transverse dc current $I_{dc}$ for the (001) (a), (110) (b) and (111) (c) oriented devices.  (d-f) $V_L^{2\omega}$ vs. $I_\omega$ at $V_g=140$ V for various values of $I_{dc}$ for the same three orientations.}      
	\label{fig:longitudinaldependenceB}
\end{figure*}  

\subsection{Sample measurement} 
The devices were measured in a Kelvinox MX100 dilution refrigerator with a base temperature of $\sim$25 mK equipped with a two-axis superconducting solenoid.  Figure \ref{fig:sampleschematic} shows a schematic of the measurement geometry.  For conventional electrical characterization, the longitudinal and transverse resistances were measured using lock-in amplifiers and a high-impedance custom current source with $\sim$100 nA of probe current, with custom voltage preamplifiers based on Texas Instrument's INA110 and Analog Devices AD624 instrumentation amplifier chips to amplify the longitudinal and transverse voltages respectively (Fig. \ref{fig:sampleschematic}).  The ac frequency used was $\sim$ 10-20 Hz.  A gate voltage $V_g$ was applied to the back of the sample substrate using a Keithley KT2400 source whose output was heavily filtered and measured independently by an Agilent 34401A multimeter.  As the properties of 2DEGs based on complex oxides are known to drift on initial cooldown, an electrostatic annealing step was performed at 4 K before any other measurement.  This consisted of cycling $V_g$ repeatedly between $\pm$200 V until the measured longitudinal and transverse differential resistances retraced.  Carrier concentrations and mobilities at different $V_g$ were also determined from conventional perpendicular field magnetoresistance measurements of the longitudinal resistance and the field antisymmetric component of the transverse resistance, based on a single band model.  The carrier concentrations and mobilities are similar to those obtained by other groups.\cite{mallik_superfluid_2022,liu_two_dimensional_2021}  In the data reported here, we restrict ourselves to $V_g>25$ V, where the sheet resistance $R_\square$ of the devices is relatively smaller ($R_\square$ rises exponentially as $V_g$ is decreased below this value).

For the nonlinear Hall effect measurements, a floating dc voltage was applied by an Agilent 33500B waveform generator to the transverse contacts as shown in Fig. \ref{fig:sampleschematic}. The voltage was passed through two 1 M$\Omega$ sourcing resistors, which were much larger than the resistance between the transverse voltage probes, effectively forming a floating passive current source that determines the transverse dc current, $I_{dc}$.  The ac current, $I_{\omega}$, was supplied through a current source using a second Agilent 33500B waveform generator with a sine wave of frequency $f\sim$17 Hz. The amplitude of the ac current drive was modulated using a function on the same waveform generator.  Four lock-in amplifiers were synchronized to the waveform generator.  Two lock-ins were tuned to measure the $f$ signal of the longitudinal $V_L$ and transverse $V_T$ voltages, and the other two were tuned to measure the $2f$ signal of $V_L$ and $V_T$.  Each lock-in was individually phased to the output of the current source at the frequency of their detection.  All data reported below were taken in zero external magnetic field. 

Since we are interested in the normal state properties, all measurements reported here were performed at $\sim$4 K, well above any superconducting transition ($\sim$ 0.2-1 K).  We shall also refer to the $f$ and $2f$ responses as the $\omega$ an $2\omega$ responses in order to conform to the published literature on the nonlinear Hall effect.

\section{Experimental Results}
\subsection{Longitudinal response}
\begin{figure*}
	\includegraphics[width=18cm]{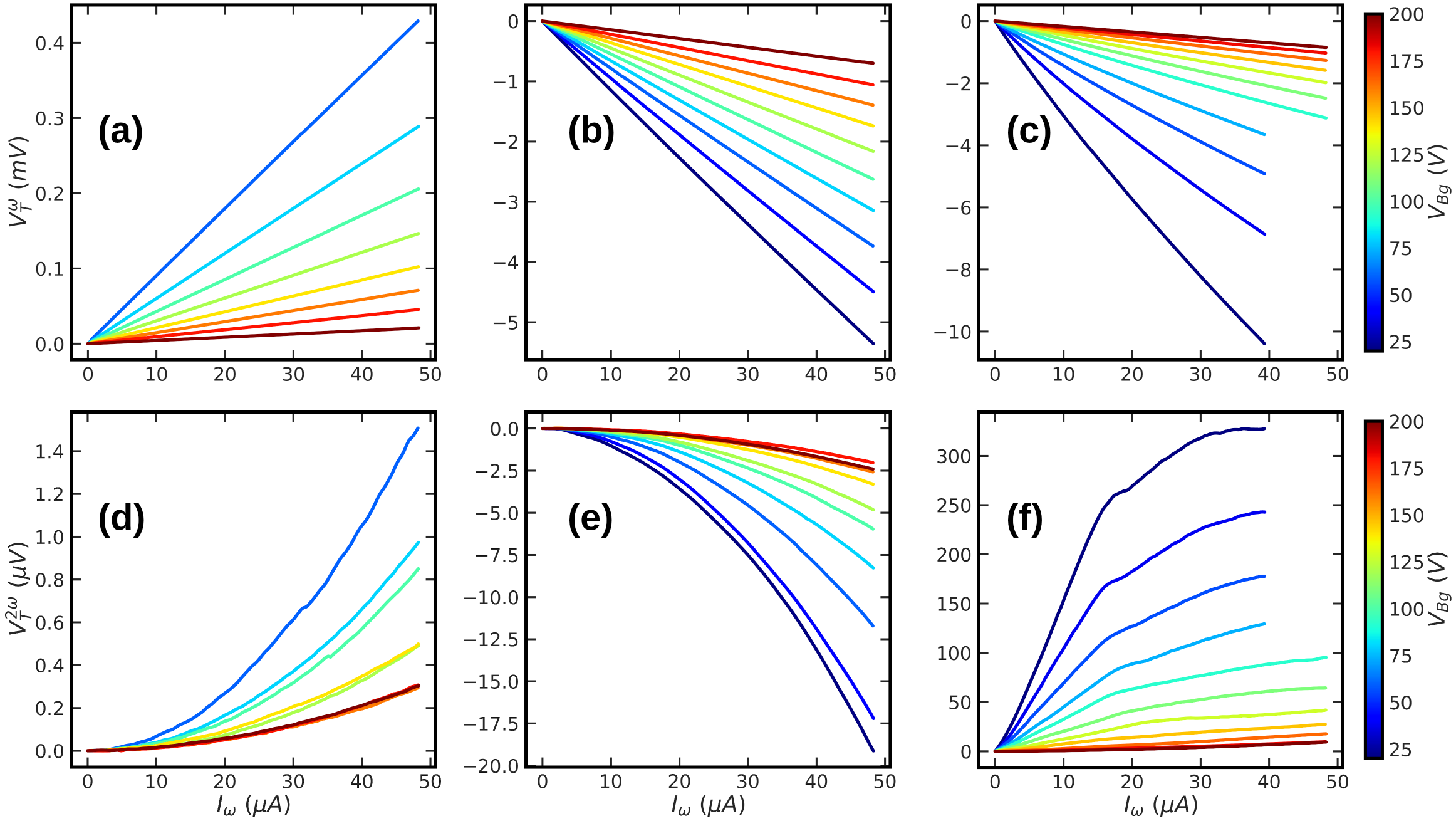}
	\caption{(a-c) Transverse voltage response at the frequency $\omega$, $V_T^\omega$, as a function of ac drive current $I_\omega$ with $I_{dc}$=0 at various values of $V_g$ for the (001) (a), (110) (b) and (111) (c) oriented devices.  (d-f) $V_T^{2\omega}$ vs. $I_\omega$ with $I_{dc}$=0 at various values of $V_g$ for the same three orientations.  There is no externally applied magnetic field.}      
	\label{fig:transverseresponse0idc}
\end{figure*}
\begin{figure*}
	\includegraphics[width=18cm]{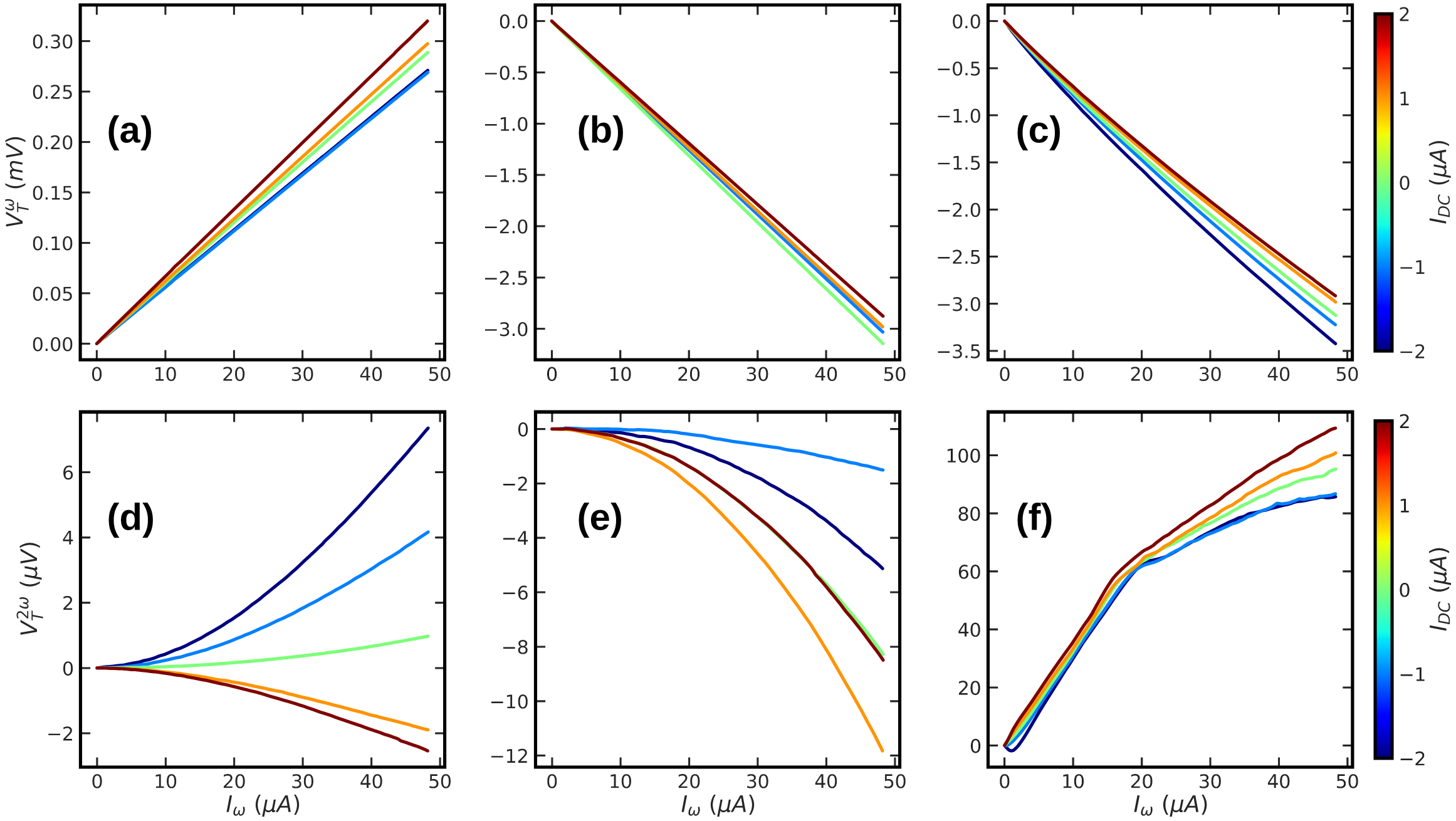}
	\caption{(a-c) Transverse voltage response at the frequency $\omega$, $V_T^\omega$, as a function of ac drive current $I_\omega$ at various values of $I_{dc}$ for the (001) (a), (110) (b) and (111) (c) oriented devices.  (d-f) $V_T^{2\omega}$ vs. $I_\omega$ for various values of $I_{dc}$ for the same three orientations.  All data taken at $V_g$ = 140 V.  There is no externally applied magnetic field.}      
	\label{fig:transverseresponse25V}
\end{figure*}  
We first discuss the $\omega$ and $2\omega$ response of the longitudinal voltage $V_L$, denoted by $V_L^\omega$ and $V_L^{2\omega}$ respectively, as a function of the ac drive current $I_\omega$ applied in the longitudinal direction, the dc current $I_{dc}$ applied in the transverse direction, and the back gate voltage $V_g$ (see Fig. \ref{fig:sampleschematic}).  Figures \ref{fig:longitudinaldependenceA}(a-c) show $V_L^\omega$ as a function of $I_\omega$ for various $V_g$ at $I_{dc}=1$ $\mu$A for the (001), (110) and (111) surface crystal orientations respectively.  $V_L^\omega$ is essentially a linear function of $I_\omega$ with the slope of the line changing as $V_g$ is varied, corresponding to the expected change in the longitudinal resistance with $V_g$.  Figures \ref{fig:longitudinaldependenceA}(d-f) show the simultaneously measured dependence of $V_L^{2\omega}$ on $I_\omega$ for the three crystal orientations under the same conditions.  $V_L^{2\omega}$ has a roughly quadratic dependence on $I_\omega$, with the curvature changing with $V_g$.  As we noted in the introduction, if the current-voltage $V_L(I)$ of the devices is not linear, one expects a finite response at higher harmonics of the drive current, with the $2\omega$ response being quadratic in $I_\omega$ for small $I_\omega$, so this behavior is as expected.  Note that in general, $V_L^{2\omega}$ is 2-3 orders of magnitude smaller than $V_L^\omega$, so the nonlinearity is small.  Similar behavior is observed for other values of $I_{dc}$.

Figures \ref{fig:longitudinaldependenceB}(a-c) show $V_L^\omega$ vs. $I_\omega$ at various values of $I_{dc}$ for $V_g=140$ V for the three different crystal orientations.  As can be seen, varying $I_{dc}$ essentially does not change the dependence of $V_L^\omega$ on $I_\omega$ if $V_g$ is kept fixed.  Figures \ref{fig:longitudinaldependenceB} (d-f) show similar data for the $V_L^{2\omega}$ response, which again does not change with $I_{dc}$.  Similar results are obtained at different fixed values of $V_g$.
  
\subsection{Transverse response}
We now discuss the transverse (Hall) response, first for the case when there is no transverse current imposed ($I_{dc}=0$).  Figures \ref{fig:transverseresponse0idc}(a-c) show the transverse ac voltage at frequency $\omega$, $V_T^\omega$, as a function of ac drive $I_\omega$ for various values of $V_g$, with $I_{dc}=0$, for all three surface crystal orientations in zero external magnetic field.  The first thing to note is that there is a finite $V_T^\omega$ in the absence of an external magnetic field for all three crystal orientations, which is smallest in the (001) device, larger in the (110) device, and largest in the (111) device.  A common source of a finite transverse signal in a Hall bar geometry in the absence of an external magnetic field is misalignment of the Hall probes due to less than perfect lithography, so that measurement of a voltage between the Hall probes includes a contribution from the longitudinal resistance.  If we calculate the ratio of the average slope of $V_T^\omega(I_\omega)$ to the slope of $V_L^\omega(I_\omega)$, it ranges from a maximum of  $\sim0.08$\% for the (001) oriented device to a maximum of $\sim1$\% for the (111) device.  Analysis of the images of the device show that the maximum possible misalignment of the Hall probes is less than 0.25 $\mu$m, so that the maximum contribution to the transverse signal due to this misalignment would be 0.25/600 or 0.04\%.  Thus, while the zero-field transverse signal of the (001) device may perhaps be explained by probe misalignment, this cannot explain the much larger transverse responses in the (110) and (111) devices.  (We discuss artifacts due to probe misalignment in more detail below.)  $V_T^\omega$ for the (001) sample is also opposite in sign $V_T^\omega$ from the (110) and (111) devices.  While the dependence of $V_{T\omega}$ on $I_\omega$ is mostly linear, there are deviations at larger values of $I_\omega$ that are most evident in the (111) sample.  

Figures \ref{fig:transverseresponse0idc}(d-f) show the corresponding $2\omega$ response $V_T^{2\omega}$.  As with the $V_T^\omega$ response, the $V_T^{2\omega}$ response increases in magnitude as we go from the (001) device to the (110) device to the (111) device.  The ratio $V_T^{2\omega}/V_T^\omega$ also increases in magnitude, i.e. the nonlinearity in $V_T$ progressively increases with crystal orientation.  In addition, while the dependence of $V_T^{2\omega}$ on $I_\omega$ is approximately quadratic for small $I_\omega$ for the (001) and (110) devices, it clearly does not have this dependence for the (111) device, even having reproducible structure as a function of $I_\omega$ that evolves smoothly with $V_g$. 

We now consider $V_T$ when $I_{dc}\neq0$.  Figure \ref{fig:transverseresponse25V} shows $V_T^\omega$ and $V_T^{2\omega}$ as a function of $I_\omega$ for $I_{dc}$ from -4 to 4 $\mu$A at $V_g=140$ V, where the variation of $V_T$ with $I_{\omega}$ is largest.  As we noted earlier, $V_L^\omega$ and $V_L^{2\omega}$ essentially do not change as $I_{dc}$ is varied, with the maximum variation being of order of 0.5-2\%  in $V_L^\omega$ and $\sim$2-4\%  in $V_T^{2\omega}$ at $I_\omega=50$ $\mu$A (Fig. \ref{fig:longitudinaldependenceB}).  $V_T^\omega$ shows a somewhat larger variation with $I_{dc}$ ($\sim$15 \%) but notably does not change sign.  In contrast, $V_T^{2\omega}$ shows large changes with $I_{dc}$, even changing sign for the (001) device, and shows nonmonotonic behavior for some values of $I_{dc}$ for the (111) device.  As with the data in Fig \ref{fig:transverseresponse0idc}, the magnitude of the signal progressively increases from the (001) device to the (110) device to the (111) device.  In particular, we note that even though the $V_T^{\omega}$ responses for the (110) and (111) devices are of the same magnitude, the $V_T^{2\omega}$ response for the (111) device is roughly an order of magnitude larger than for the (110) device.         
   
\section{Discussion}

\subsection{Potential experimental artifacts}
The data shown above demonstrate a nonlinear Hall effect, and in many respects are similar to data reported on other systems as evidence of Berry dipole effects \cite{zhang_high_mobility_2018}.  However, before we consider potential topological origins of our results, it is important to consider non-topological explanations.  We consider two potential experimental artifacts below.

\subsubsection{Probe misalignment}
As we noted earlier, lithographically patterned Hall bar geometries may have a misalignment of the Hall probes, leading to a longitudinal component of resistance being added to any transverse component.  Measurements of the photomask used to fabricate our devices as well as measurements of the devices restrict this to be less than 0.04\% of the longitudinal response.  Thus the magnitude of the transverse response, particularly for the (110) and (111) oriented samples cannot arise solely from probe misalignment.  We show below that the dependence of $V_T^{2\omega}$ on $I_{ac}$ and the transverse dc current $I_{dc}$ is also inconsistent with this explanation.

As we noted above, if the IV characteristic of the device is nonlinear, one might expect a 2$\omega$ contribution to the longitudinal resistance that scales as $I_\omega^2$.  For a system with small nonlinearities, as is the case with our devices, one can expand the voltage $V$ across as a function of the current $I$ as
\begin{equation}
V(I) = \alpha I + \beta I^2 + \gamma I^3 + \dots +
\label{eqn:IVexpansion}
\end{equation}   
to the third power in $I$.  Now, one might expect that $V$ should be an antisymmetric function of $I$, so that the coefficients of the even powers of $I$ in this expansion vanish.  However, as we shall see below, this would mean that there would be no $2\omega$ response, so we include it here.  (One potential source for terms even in the current is the Seebeck effect, which is known to give rise to terms that go as $I^2$.)  If we now consider a sinusoidal current of the form $I_\omega \sin(\omega t)$, we have
\begin{align}
\label{eqn:withnodc}
V(I_\omega \sin \omega t) = &\frac{\beta}{2} I_\omega^2 + \left(\alpha  + \frac{3}{4} \gamma  I_\omega^2\right) I_\omega \sin \omega t \\ \nonumber
&- \frac{\beta}{2}I_\omega^2 \cos 2\omega t  - \frac{1}{4} \gamma I_\omega^3 \sin 3 \omega t + + . 
\end{align}
There will be corrections to this expression from higher order terms in Eqn. (\ref{eqn:IVexpansion}) which we can ignore if the nonlinearity in the IV is small.  Consequently, the amplitude of the $1\omega$ signal scales with $I_\omega$ as $(\alpha I_\omega + (3/4)\gamma I_\omega^3)$, while the $2\omega$ response (with the correct phasing) goes as $-(\beta/2)I_\omega^2$.  Applying this analysis to the data shown in Figs. \ref{fig:longitudinaldependenceA} and \ref{fig:longitudinaldependenceB}, and restricting ourselves to the region around $I_\omega=0$ where the expansion is valid, we see that $\beta>0$ for all three orientations.  In addition, fitting $V_L^\omega(I_\omega)$ around $I_\omega=0$, we find that $\gamma<0$, which agrees with other measurements that show that the resistance of the devices decreases with increasing dc current in this temperature range.  Typically, we find that $\beta$ is approximately 3-4 orders of magnitude smaller than $\alpha$, and $\gamma$ is 5 orders of magnitude smaller than $\alpha$.  Importantly for what we discuss below, $\beta$ (in units of mV/($\mu$A)$^2$) is typically approximately a factor of 10 greater than $\gamma$ (in units of mV/($\mu$A)$^3$). 

For completeness, we consider the case when there is a dc current $I_{dc}$ in addition to the ac current.  Expanding $V(I_{dc}+I_\omega \sin \omega t)$, one finds that the $1\omega$ and $2\omega$ responses are
 \begin{subequations}
\begin{align}
1 \omega&: \quad \left(\alpha + \frac{3}{4} \gamma I_\omega^2 + 2 \beta I_{dc} + 3 \gamma I_{dc}^2\right)  I_\omega \sin \omega t \\ \label{eqn:withdca}
2 \omega &: \quad - \frac{1}{2}(\beta + 3 \gamma I_{dc})I_\omega^2 \cos 2\omega t 
\end{align}	 	
\end{subequations} 	 
which reduce to the equivalent terms in Eqn. \ref{eqn:withnodc} when $I_{dc}=0$.  Since $\alpha>>\beta,\gamma$, we expect that applying a dc current in addition to an ac current will only slightly modify the slope of $V^\omega(I_\omega)$.  However, since $\beta$ is only a factor of 10 larger than $\gamma$, a dc current of magnitude of a few $\mu$A's can change the sign of the curvature of $V^{2\omega}(I_\omega)$, depending on the sign of $I_{dc}$.

\begin{table*}[htb]
	\begin{tabular}{| c | c | c | c | c |}
		\toprule
		\textbf{ Orientation} & $\alpha$ (mV/$\mu$A) & $\beta$ (mV/($\mu$A)$^2$ $\times 10^{-4})$ & $\gamma$ (mV/($\mu$A)$^3$ $\times 10^{-5}$) & $\beta/3\gamma$ ($\mu$A)\\
		\midrule
		(001) & 9.02 & 5.58 & -8.13 & -2.3 \\
		\hline
		(110) & 7.51  & 3.42 & -25.96 & -0.3 \\
		\hline
		(111) & 12.63  & 22.62 & -26.57 & -2.9 \\
		\bottomrule	
	\end{tabular}
	\caption{Parameters for fit to the $1\omega$ and $2\omega$ contribution of Eqn. (\ref{eqn:withdca}) of the $V_g=80 $V longitudinal response shown in Fig. \ref{fig:longitudinaldependenceA} with $I_{dc}=0$.  Fits were confined to $0 \leq I_\omega \leq$ 20 $\mu$A.}
	\label{tab:fitparameters}
\end{table*}

With this analysis in mind, let us return to the data shown in Figs. \ref{fig:transverseresponse0idc} and \ref{fig:transverseresponse25V}, initially considering the possibility that the finite transverse voltage observed is due to a misalignment of the transverse probes, so that transverse voltage we observe is due to a small longitudinal component between the measurement probes.  If this is the case, the dependence of $V_T^{\omega}$ and $V_T^{2\omega}$ on $I_\omega$ should reflect the dependence of $V_L^\omega$ and $V_L^{2\omega}$ on $I_\omega$.  We focus first on the (001) sample, which as we noted above is the one that has the smallest zero-field transverse voltage, and hence is the sample most likely to fit this scenario.  Comparing Figs. \ref{fig:longitudinaldependenceA}(a) and (d) to Figs, \ref{fig:transverseresponse0idc}(a) and (d), we see immediately that while the slopes of $V_L^\omega$ and $V_T^\omega$ have the same sign, the curvatures of $V_L^{2\omega}$ and $V_T^{2\omega}$ are opposite in sign, the first indication that the transverse signal does not arise from a simple misalignment of the probes, even for the (001) sample.  Although the signs are reversed, similar behavior is seen for the (110) device, in which the transverse response is also larger in magnitude.  For the (111) device, the overall dependence of $V_T$ on $I_\omega$ is consistent with the behavior of $V_L$ on $I_\omega$.

Further evidence that the transverse signals we observe are not due to a misalignment of the Hall probes can be obtained by a more detailed numerical analysis of the data shown in Fig. \ref{fig:transverseresponse25V}.  To facilitate this analysis, Table \ref{tab:fitparameters} shows the values of $\alpha$, $\beta$ and $\gamma$ of Eqn. (\ref{eqn:IVexpansion}) obtained by fitting $V_L^\omega(I_\omega)$ and $V_L^{2\omega}(I_\omega)$ of the $V_g=80$ V data of Fig. \ref{fig:longitudinaldependenceA} to the $1\omega$ and $2\omega$ responses shown in Eqn. (\ref{eqn:withnodc}), restricting the range of the fit to $0\leq I_\omega \leq 20$ $\mu$A.  The parameters so obtained should then enable us to predict the $I_{dc}$ dependence of $V_T^{2\omega}(I_\omega)$ shown in Fig. \ref{fig:transverseresponse25V}.  Examining Eqn. (\ref{eqn:withdca}), we see that the curvature of $V_T^{2\omega}(I_\omega)$ depends on the ratio $(3\gamma I_{dc}/\beta)$.  Thus, when $|I_{dc}|>|\beta/3\gamma|$, the curvature of $V_T^{2\omega}(I_\omega)$ should change sign.  The last column in Table \ref{tab:fitparameters} shows the ratio $\beta/3 \gamma$ for the three crystal orientations.  Since this ratio is negative, the change in curvature should occur for positive values of $I_{dc}$.  

Consider first the data for the (001) oriented device, shown in Fig. \ref{fig:transverseresponse25V}(d).  As we noted above, if the transverse response is due to a longitudinal component arising from misalignment of the Hall probes, $I_{dc}$ would add to $I_{ac}$ in the short section where the probes are misaligned.  One also expects to see a negative curvature in $V_T^{2\omega}(I_\omega)$ from Eqn. \ref{eqn:withnodc}, since $\beta$ is positive.  Instead, we see a slight positive curvature for $I_{dc}=0$.  However, the expected dependence on $I_{dc}$ discussed above does appear to be borne out, in that the curvature becomes increasingly positive for $I_{dc}<0$, and increasingly negative for $I_{dc}>0$, indeed changing sign for the largest values of $I_{dc}$.

$V_T^\omega$ for the (110) device (Fig. \ref{fig:transverseresponse25V}(b)) has the opposite sign in comparison to the (001) device, so if we wish to ascribe the finite transverse voltage to a misalignment of the probes and compare it to the data from the (001) device, we must reverse the signs of both $V_T^{\omega}$ and $V_T^{2\omega}$.  Keeping this is mind,  $V_T^{2\omega}$ for the (110) oriented device (Fig. \ref{fig:transverseresponse25V}(e)) also has the wrong curvature, but with a much larger response at $I_{dc}=0$.  Consequently, following our analysis above, with the data as plotted in Fig. \ref{fig:transverseresponse25V}(e), an increasingly positive $I_{dc}$ should give rise to an increasing curvature.  However, the opposite trend is observed: increasing $I_{dc}$ decreases the curvature.  Furthermore, the dependence of the curvature of $V_T^{2\omega}$ on $I_{dc}$ is non-monotonic, which clearly does not fit into the analysis presented above. 

Finally, for the (111) oriented sample, while the relative signs of $V_T^{\omega}$ and $V_T^{2\omega}$ agree with those of $V_L^{\omega}$ and $V_L^{2\omega}$, the dependence of $V_T^{2\omega}$ on $I_\omega$ is clearly not quadratic, showing roughly linear behavior up until around $I_{ac}\sim 15-20$ $\mu$A, followed by a kink and a transition to a more complicated dependence.  The dependence on $I_{dc}$ is also relatively much weaker than in the (001) and (110) samples.  

In summary, while there might be a small contribution to the zero field transverse signal we observe due to probe misalignment, both the magnitude of the signal as well as its dependence on $I_\omega$ and $I_{dc}$ show that the major contribution does not arise from probe misalignment, particularly in the (110) and (111) oriented devices.

More generally, the expansion that we performed for the longitudinal voltage (Eqn. (\ref{eqn:IVexpansion})) should also apply to the transverse voltage $V_T(I)$ regardless of its origin.  If the $2\omega$ response of the transverse voltage is due simply to the nonlinear dependence of $V_T(I)$ on $I$, then the $2\omega$ response should show a quadratic dependence on $I_\omega$ for small $I_\omega$.  While this appears to be the case for the (001) and (110) orientations, the sign of $\beta$ is different in comparison to the longitudinal response, while for the (111) orientation the dependence is clearly not quadratic.  In addition, the magnitude of the $2\omega$ response is larger for the transverse signal in comparison to the longitudinal response, particularly for the (110) and (111) orientations, suggesting a different origin for the $2\omega$ response of the transverse signal. 
 
\subsubsection{Frequency dependence}
\begin{figure*}
	\includegraphics[width=18cm]{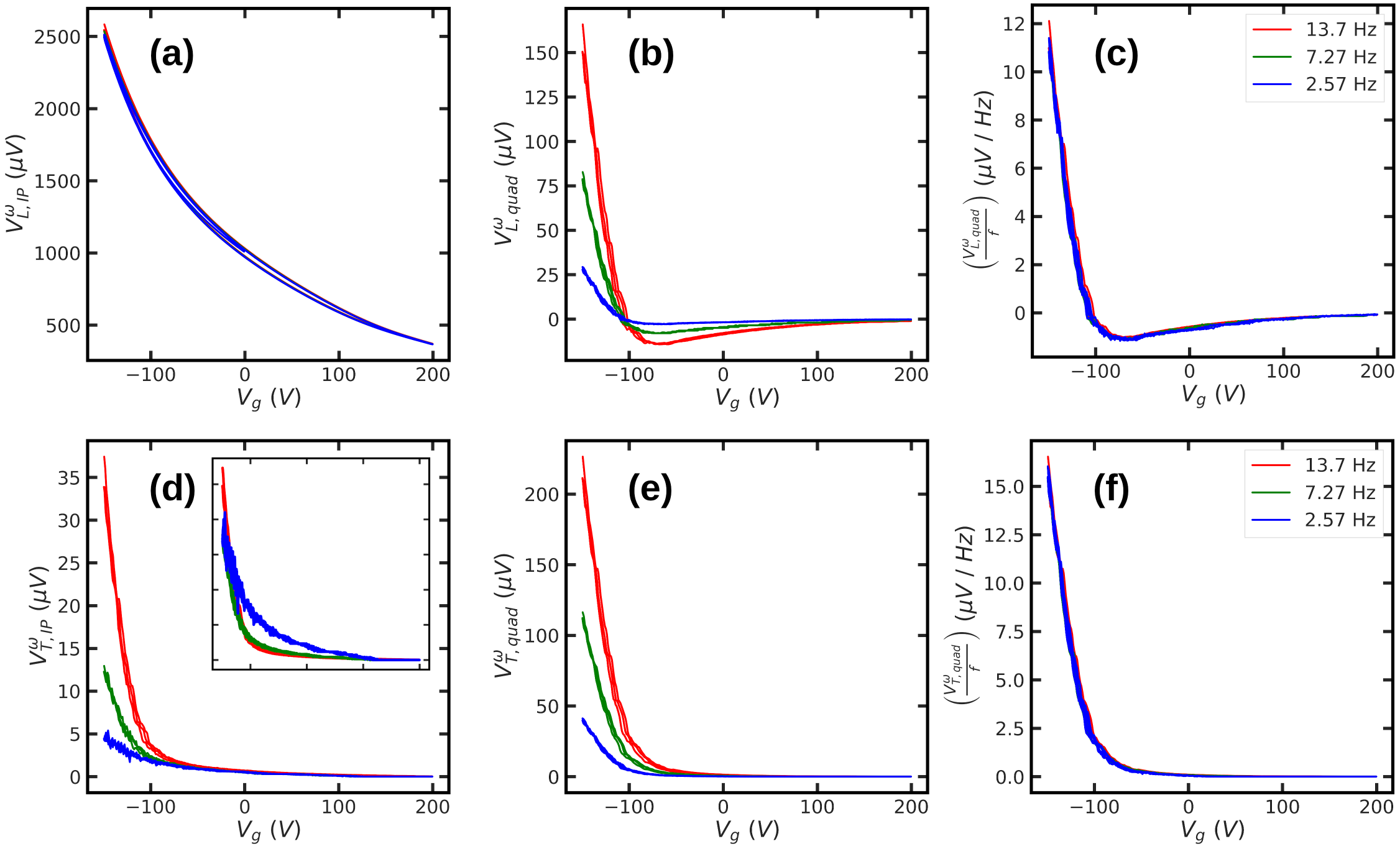}
	\caption{(a)  In-phase longitudinal voltage response of the (001) device as a function of $V_g$ at three different frequencies of the ac current drive of 100 nA.  (b)  Simultaneously measured quadrature longitudinal response.  (c)  Quadrature response in (b) divided by the measurement frequency.  (d)  In-phase transverse response.  (e)  Simultaneously measured quadrature transverse response.  (f)  Quadrature response divided by ac measurement frequency.  The inset to (f) shows the transverse in-phase response divided by the ac measurement frequency.}      
	\label{fig:frequencydependence}
\end{figure*}  
A second, more subtle potential experimental artifact arises from the highly resistive nature of the devices, particularly at lower values of $V_g$ where the charge density is lower and the 2DEGs are presumably more disordered.  Since the current and voltage contact probes of the Hall bar also become more resistive, it is important to use measurement techniques that take this into account.  To this end, we use custom built current sources with output impedances in excess of $10^{12}\;\Omega$, and input preamplifiers (Texas Instruments INA110) with input impedances of $10^{12}\;\Omega$ in order to enable continuous measurements where the resistance of the devices may vary by orders of magnitude.  Nevertheless, the highly resistive nature of the samples at lower $V_g$, coupled with the capacitances of the measurement lines and the not-insignificant capacitance of the 2DEG to the conducting back gate may affect the measurement.  Evidence of this has been observed in LAO/STO devices\cite{davis_anisotropic_2017} and LSAT/STO devices at large negative $V_g$.\cite{bal_electrostatic_2017}

\begin{figure}
	\includegraphics[width=8cm]{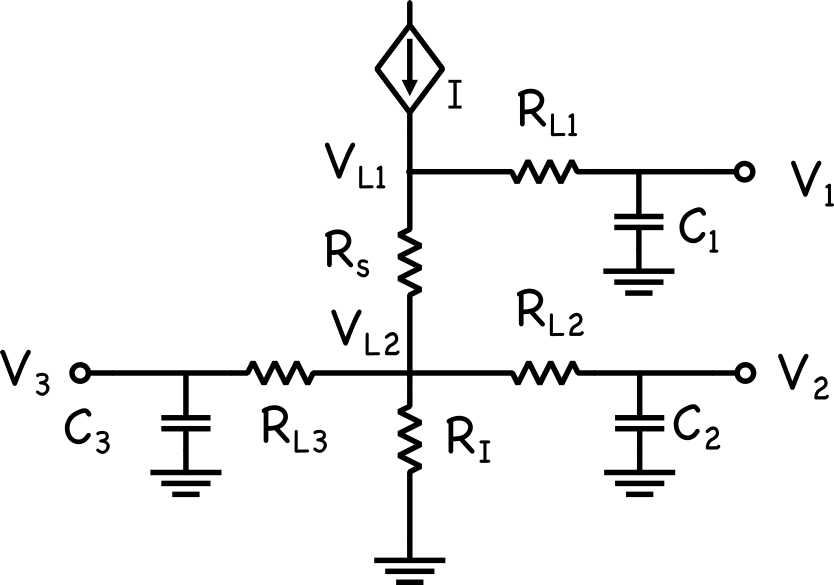}
	\caption{Schematic of Hall bar used to model frequency dependence}
	\label{fig:freqdependenceschematic}
\end{figure}
To understand how these factors may affect the measurement, we have measured the in-phase and quadrature signals of the longitudinal and transverse response for three different measurement frequencies.  In-phase here signifies the signal that is in-phase with the ac current drive, monitored by measuring the voltage across a 10 k$\Omega$ sense resistor placed in series with the sample; the quadrature response is the signal that is 90$^\circ$ out of phase with the voltage across the sense resistor.  Figures \ref{fig:frequencydependence}(a) and (b) show the in-phase and quadrature ac voltage at the drive frequency $V_{L,ip}^\omega$ and $V_{L,quad}^\omega$  as a function of $V_g$ at three different frequencies $f$ of the ac current drive, using a drive amplitude of 100 nA.  These data are for the (001) device.  The in-phase response is essentially independent of $f$ over this frequency range.  However, the quadrature response, shown in Fig. \ref{fig:frequencydependence}(b), clearly depends on frequency with the response being smaller for lower values of frequency.  Dividing the quadrature response by $f$, all three curves collapse on to a single curve (Fig. \ref{fig:frequencydependence}(c)), showing that the quadrature response scales with $f$, and hence appears to be purely reactive.            
   
Figures \ref{fig:frequencydependence}(d-f) show similar data for the transverse voltage $V_T^\omega$.  Here, the in-phase response (Fig. \ref{fig:frequencydependence}(d)) is not independent of $f$, with a clear difference visible below $\sim V_g\leq -75$V, being smaller for lower values of $f$.  The quadrature response (Fig. \ref{fig:frequencydependence}(e)) also decreases with $f$, but dividing by $f$ as we did for the longitudinal response, we find that all the curves again collapse on to a single curve (Fig. \ref{fig:frequencydependence}(f)), indicating again that the quadrature response is purely reactive.  Performing a similar frequency scaling to the in-phase component does not result in a single curve (inset to Fig. \ref{fig:frequencydependence}(d)).  

To attempt to model this behavior, we consider a simple model of a Hall bar device including the resistance and capacitance in each contact.  These are likely to be distributed resistances and capacitances, but we model them as single reduced elements, as shown in Fig. \ref{fig:freqdependenceschematic}.  (Similar analysis for LAO/STO structures show both models give similar results \cite{davis_anisotropic_2017}.) Here $R_s$ is the nominal Hall bar longitudinal resistance, $R_{L1}$, $R_{L2}$, $R_{L3}$ and $R_I$ are the resistances in the contact leads, and $C_1$, $C_2$ and $C_3$ are the capacitances in the leads (we ignore the capacitances in the current contacts, and the resistance in the $I^+$  lead as they do not enter in the analysis).  Let us consider first the longitudinal response, and to further simplify the analysis, let us assume that $R_{L1}=R_{L2}=R_L$, and $C_1=C_2=C$.  Then it is easy to show that measured voltage drop $V_1-V_2$ is related to the voltage drop $V_{L1}-V_{L2}$ across $R_s$ by
\begin{equation}
(V_1-V_2) = (V_{L1}-V_{L2}) \frac{1-j \omega R_L C}{1 + \omega^2 R_L^2 C^2}
\end{equation}
where we have used the engineering notation of $j=\sqrt{-1}$.  This indicates that the ratio of the measured quadrature signal to the measured in-phase signal is $\omega R_L C$.  An estimate of $R_L C$ can be obtained by comparing the in-phase and quadrature signals shown in Figs. \ref{fig:frequencydependence}(a) and (b).  For example, if we compare the magnitudes of the in-phase and quadrature signals for $f=13.7$ Hz at $V_g=-50$ V, we get $R_L C \sim$ 0.1 ms, which suggests that the effects of the lead resistances/capacitances are small even down to this gate voltage.

Let us now consider the measured transverse response $V_3-V_2$. In the absence of a zero-field transverse signal and any misalignment of the Hall probes, $V_3=V_2$ if $R_{L3}=R_{L2}$ and $C_3=C_2$ as we assumed for the longitudinal case, so that the transverse response vanishes.  If we do have a misalignment of the Hall probes, then the analysis that we applied for the longitudinal case should also apply here...in particular, making the reasonable assumption that the resistances and capacitances each contact are of the same order, we should expect a quadrature component that is much smaller than the in-phase component. In contrast, we observe a quadrature component that is almost an order of magnitude larger than the in-phase component.  This observation is another indication that the transverse component we observe is not solely due to probe misalignment.

A second scenario is that the true zero field transverse resistance vanishes, but differences between the resistances and capacitances in the two Hall leads result in a finite measured difference $V_3-V_2$.  A simple analysis equivalent to the longitudinal case gives
\begin{equation}
V_3-V_2 = V_{L2} \left( \frac{1 - j \omega R_{L3}C_3}{1+ \omega^2 R_{L3}^2 C_3^2} -\frac{1 - j \omega R_{L2}C_2}{1+ \omega^2 R_{L2}^2 C_2^2}\right).
\end{equation} 
At low frequencies, this gives
\begin{align}
V_3-V_2 \simeq  - V_{L2}  &\left[ \omega^2 (R_{L3}^2 C_3^2 - R_{L2}^2 C_2^2)  \right. \nonumber \\
& \left. - j \omega (R_{L3}C_3 - R_{L2}C_2) \right]
\end{align}  

From our analysis of the longitudinal response, $\omega R_L C \sim 10^{-2}$, so with this scenario, the in-phase component of the transverse response should be approximately this factor smaller than the quadrature component.  The data of Fig. \ref{fig:frequencydependence} show that the in-phase signal is about 17\% of the quadrature signal at the maximum negative gate voltage $V_g \sim -150$ V, even for the (001) oriented device, which has the smallest transverse response.  At larger $V_g$ ($>0$ V), the quadrature component essentially vanishes while the in-phase response is still finite.  Comparing the expected quadrature longitudinal and transverse signals, both are proportional to $\omega R_L C$.  The longitudinal response is proportional to $(V_{L1}- V_{L2)})$, while the transverse response is proportional to $V_{L2}$.  Comparison of 2-terminal and 4-terminal resistances show that $V_{L2}$ is within a factor of 2 of $V_{L2}-V_{L1}$, so that the quadrature components of the longitudinal responses should be comparable, consistent with our observations.  However, the in-phase response is much larger than expected from this analysis.  Again, it should be emphasized that the data in Figs. \ref{fig:frequencydependence}(d-f) is for the (001) sample, whose transverse response is the smallest. As the transverse quadrature response should be about the same for all the orientations as the sample resistances and capacitances are comparable, we conclude that the transverse signal that we observe is not due to the finite capacitance of the devices, particularly in the gate voltage regime $V_g>25$ V.  We thus consider the possibility that the second harmonic response arises from Berry curvature effects.

\subsection{Berry dipole effects}
\begin{figure*}[htb]
	\includegraphics[width=16cm]{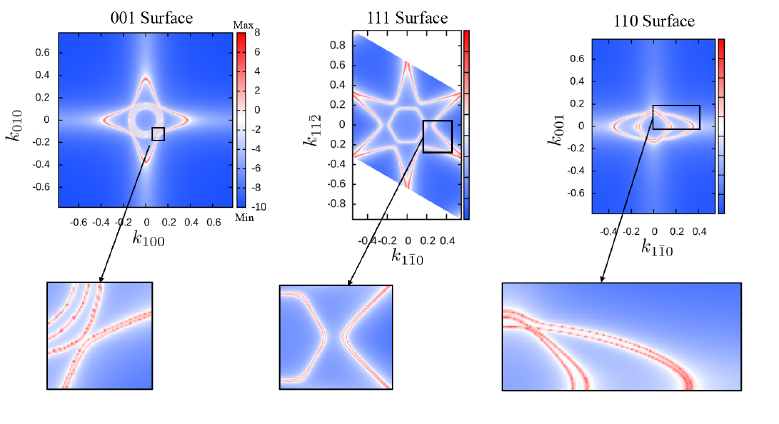}
	\caption{Band structure of the (001), (111) and (110) KTO surfaces.  Lower panels show zoomed-in regions.}      
	\label{fig:ktosurfacestates}
\end{figure*} 

\begin{figure*}[htb]
	\includegraphics[width=16cm]{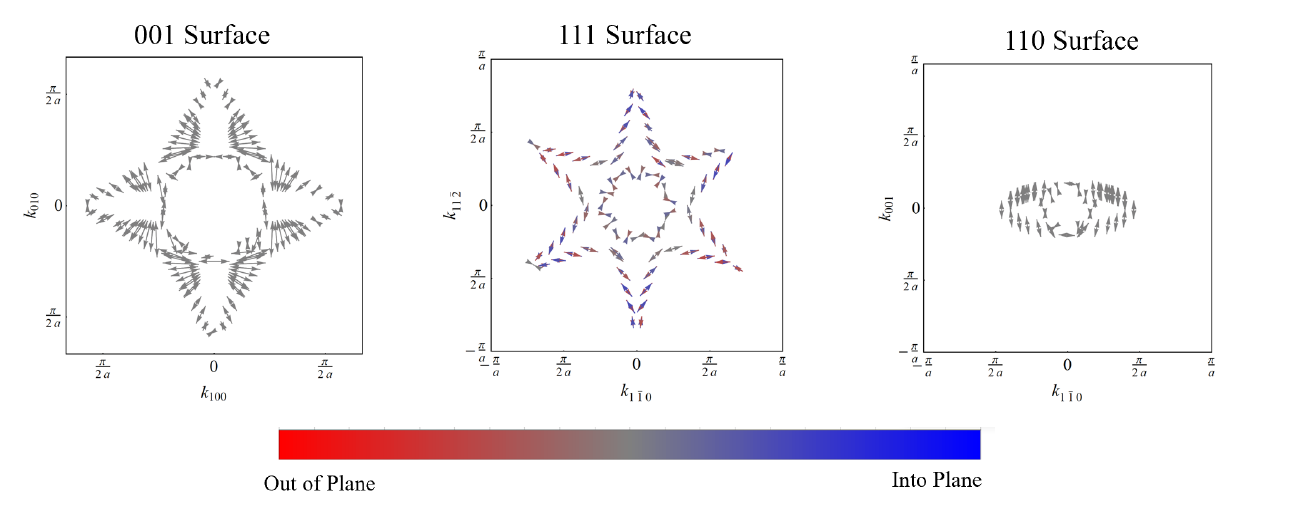}
	\caption{Spin texture of the (001), (111) and (110) KTO surface bands.  The color coding shows the out-of-plane component of the spin.}      
	\label{fig:ktospintexture}
\end{figure*}

As with its sister complex oxide SrTiO$_3$ (STO), the conduction bands at the surface in KTO are determined by the $d$-orbital $t_{2g}$ manifold. In the case of STO, these are derived from the Ti $3d$ orbitals, while in the case of KTO, they arise from the Ta $5d$ orbitals.  As a consequence of the larger atomic number of Ta in comparison with Ti, the atomic spin-orbit interaction in KTO is more than a factor of 20 larger than in STO \cite{bruno_band_2019}.  In both STO and KTO, the surface breaks the bulk inversion symmetry, leading to a Rashba spin-splitting of the bands that depends on the surface crystal orientation.  The stronger spin-orbit interactions in KTO lead to a rich spin texture of the conduction bands that has been discussed earlier specifically for the case of the (111) oriented surface \cite{bruno_band_2019}.  As with STO, the spin-splitting depends on the $k$-space direction.  Figure \ref{fig:ktosurfacestates} shows the surface bands for the (001), (110) and (111) KTO surfaces calculated using Density Functional Theory (DFT) (details of the calculation can be found in the Appendix).  Figure \ref{fig:ktospintexture} shows the corresponding spin-texture of the bands, color-coded to denote the out-of-plane component of the spins.  As noted by Bruno \textit{et al.} \cite{bruno_band_2019}, the spins in the (111) surface cant out of the plane of the 2DEG.  For the (001) and (110) orientations, we find that the spins remain in the plane of the 2DEG.

\begin{figure*}[ht]
	\includegraphics[width=16cm]{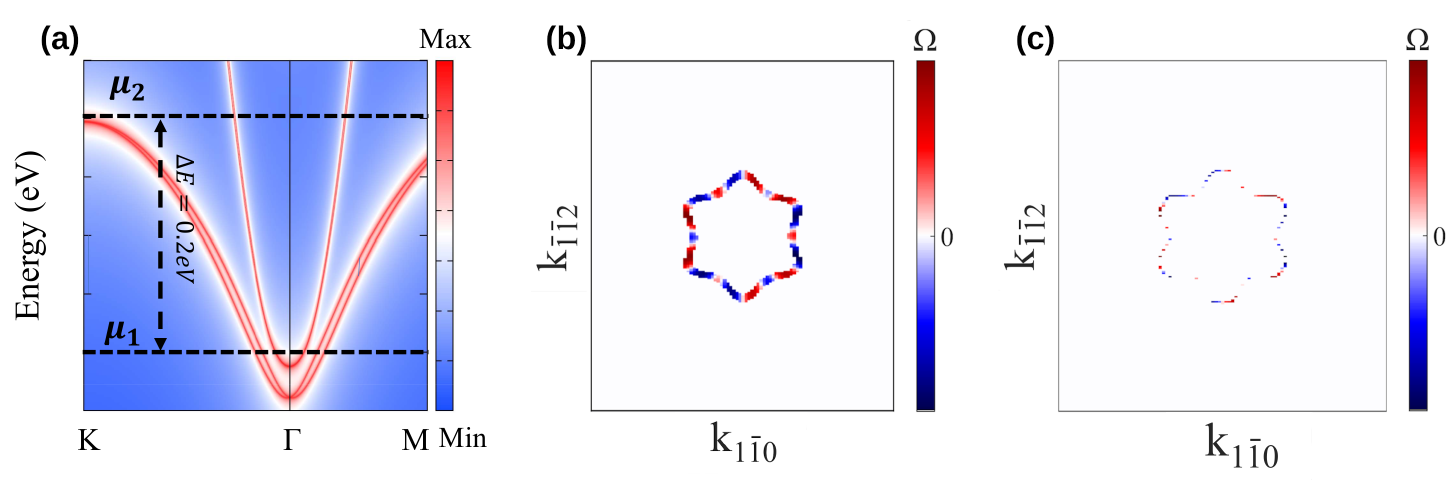}
	\caption{(a)  Band dispersion of the (111) surface.  (b,c) Calculated Berry curvature at the chemical potential $\mu_1$ (b) and $\mu_2$ (c) shown in (a).}      
	\label{fig:berrycurvature}
\end{figure*}

Figure \ref{fig:berrycurvature} shows the calculated Berry curvature for the (111) surface orientation at two different values of the chemical potential $\mu$.  Both the (001) and (110) surfaces are inversion symmetric so that the Berry curvature identically vanishes in the presence of time-reversal symmetry, a result that is supported by our DFT calculations.  For the (111) surface, a finite Berry curvature appears in those regions of $k$ space where only one band of a pair of spin-split bands is occupied for a specific value of $\mu$, since the Berry curvatures of the two spin bands are opposite in sign and cancel each other if both are occupied.  We note that in general, the Berry curvature is larger for smaller values of $\mu$, corresponding to lower values of $V_g$.

From the general symmetries of the three surfaces and the DFT calculations presented above, one should not expect a nonlinear Hall response from any of the three surfaces in the absence of a transverse electric field (or current). In the presence of such a transverse field, one might expect a contribution from the (111) surface as discussed in the introduction, and perhaps also from the (001) and (110) surfaces if we think of the transverse field as breaking the inversion symmetry in these surface orientations.  Nevertheless, as we have shown above, we observe a clear transverse nonlinear Hall responses in all three orientations, with the magnitude progressively increasing as we go from the (001) to the (110) to the (111) surface.  

What is the possible origin of the nonlinear Hall effect that we observe?  We discuss several possibilities.  The first is that relaxation or strain at the surface reduces its symmetry, loosening some of the constraints that force the Berry curvature to vanish for the (001) and (110) surfaces, and eliminating some of the mirror lines for the (111) surface and thereby allowing a finite Berry dipole moment.  One could also imagine miscuts of the crystal substrates, typically of the order of a few degrees, might lead to similar effects.  However, such effects might be expected to give nonlinear responses of the same order, particularly for the (001) and (110) surfaces, while experimentally we observe that the response in the (110) surface is approximately an order of magnitude larger than in the (001) surface, with the (111) surface showing an even greater response.  A second possibility is that intrinsic magnetism in the system breaks time-reversal symmetry, leading to a finite transverse response.  There have been reports of intrinsic magnetism in KTO 2DEGs, and we have seen evidence of long range magnetic order at low temperatures in the present devices.  This might also explain the finite transverse $1\omega$ response we see in all the samples.  However, in a system with inversion symmetry but broken time-reversal symmetry, the Berry curvature is an even function of the momentum, $\boldsymbol{\Omega}(\boldsymbol{k}) = \boldsymbol{\Omega}(\boldsymbol{-k})$, so that the Berry dipole would vanish, and one would not expect to see a second harmonic signal in the transverse response.  A third possibility is that the finite thickness of the 2D layer means that higher bands are likely occupied, as is known from the case of STO.  The Berry phase calculations that we have performed are for $k_z = 0$:  since inversion symmetry in the $z$ direction is broken by the surface, bands with finite $k_z$ might give rise to Berry dipoles with components in the plane of the 2DEG.  This possibility has not yet been discussed theoretically in the literature to our knowledge.   

\section{Summary}
In summary, we have performed transport measurements in KTO 2DEGs of three different surface orientations, (001), (110) and (111).  All three surface orientations show a nonlinear Hall effect, i.e., a finite response in the second harmonic of the transverse voltage at frequency $2f$ in response to an applied longitudinal ac drive current at frequency $f$.  The nonlinear Hall response progressively increases in magnitude in going from the (001) to the (110) surface orientation, depend on the applied back gate voltage, and can be modified by a transverse dc current.  This trend suggests the nonlinear Hall effect might be related to Berry curvature effects as DFT calculations show that these are expected to be strongest in the (111) oriented devices, but further work is required to clarify the relation between the potential topological nature of the band structure and our experimental observations. 

\section*{Appendix: Computational Methods}
\paragraph*{Surface 2DEG and Spin Texture}
KTaO$_{3}$ belongs to spacegroup 221 with lattice parameter $a=4.03$ \AA. The lattice parameters and atomic positions are taken from the Materials Project \cite{jain_commentary_2013}. All first principles calculations are based on density functional theory and were performed using the Quantum ESPRESSO software package \cite{giannozzi_q_2020}. The calculations utilize the generalized gradient approximations (GGA) of Perdew-Burke-Ernzerhoff (PBE) \cite{perdew_generalized_1996}. Spin-orbit coupling is included in the calculations.

\begin{figure*}
\centering
\subfigure[]{
\includegraphics[scale=0.4]{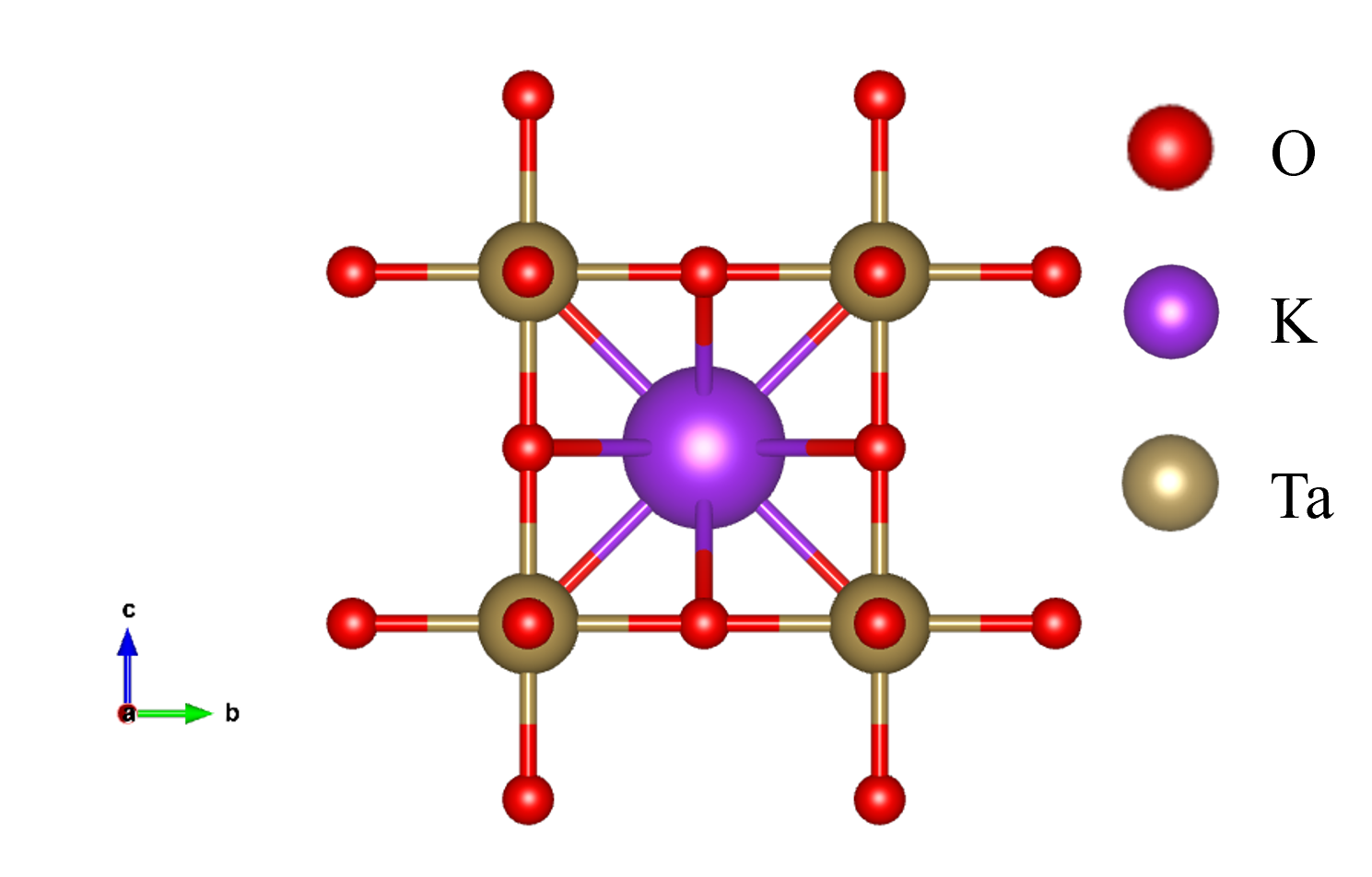}
\label{fig:structure}}
\subfigure[]{
\includegraphics[scale=0.4]{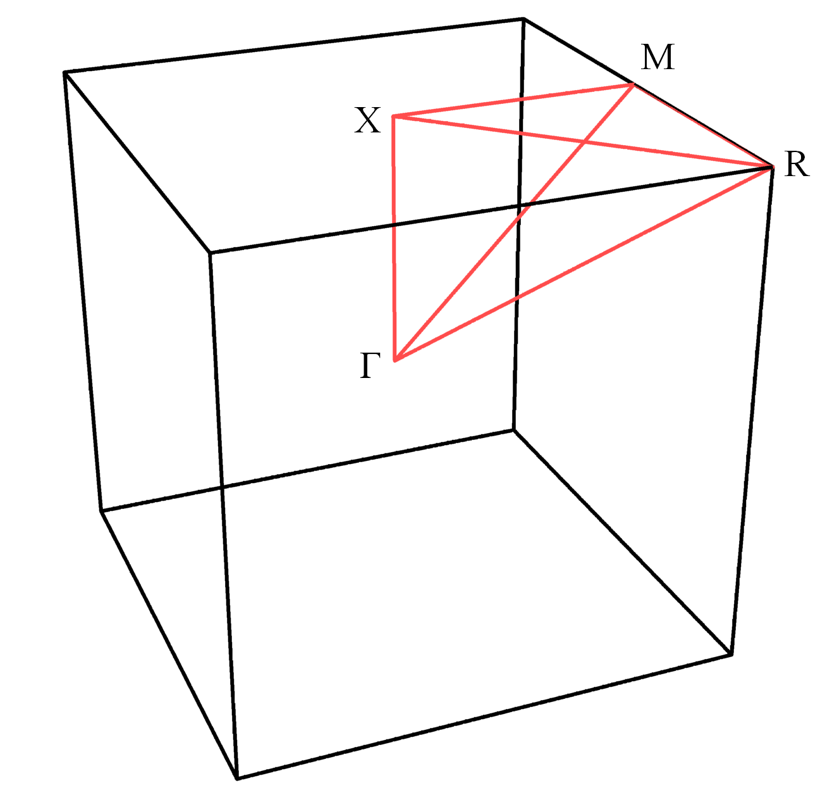}
\label{fig:Brillouinzone}}
\subfigure[]{
\includegraphics[scale=0.35]{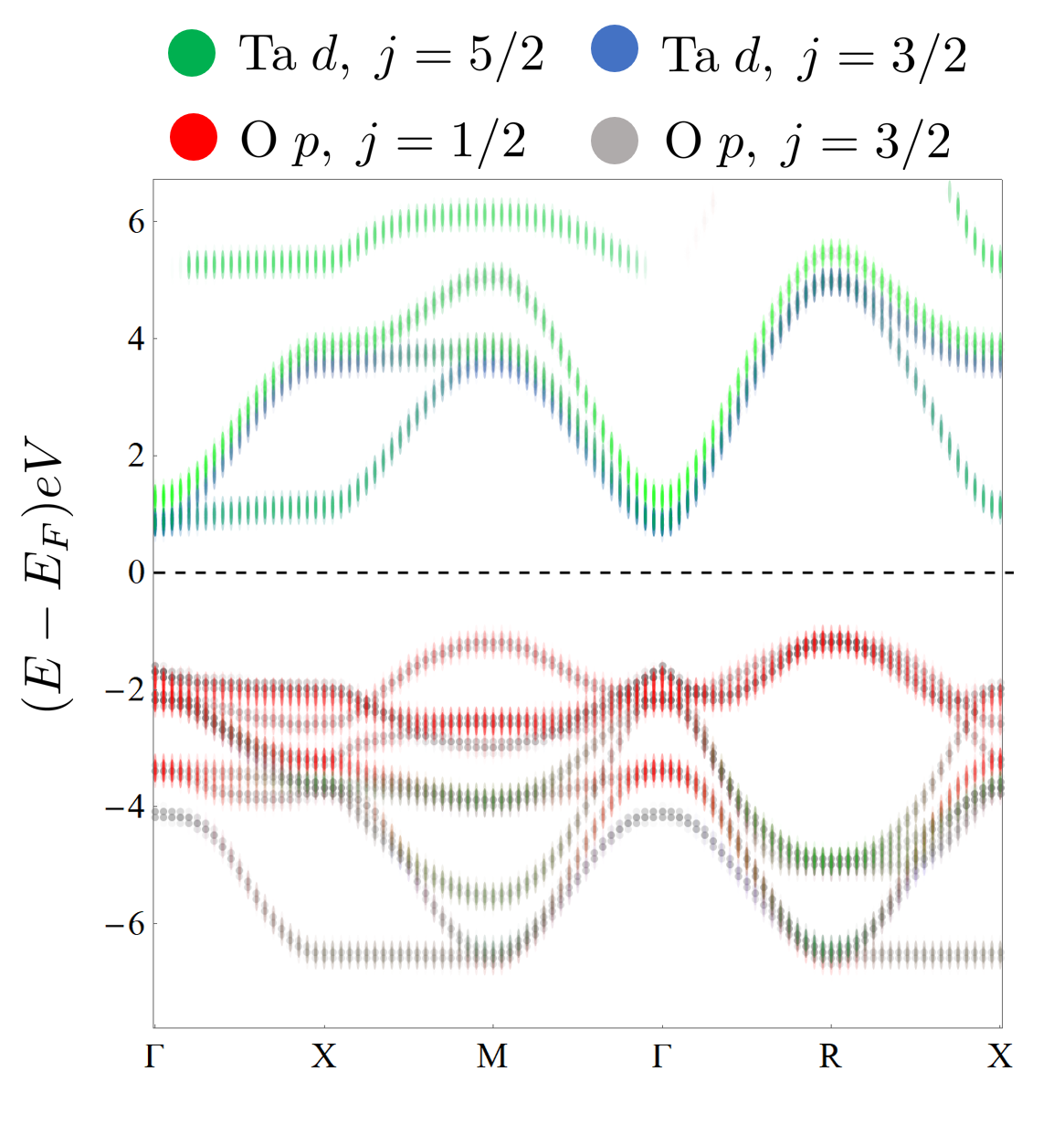}
\label{fig:Bands}}
\caption{(a) Primitive unit cell of bulk KTaO$_{3}$ belonging to spacegroup 221. (b) Brillouin zone of primitive unit cell with high-symmetry locations and path labeled. (c) Momentum space resolved density of states along the high-symmetry path labeled in (b), demonstrating that the conduction bands are primarily composed of the Ta $d$-orbitals.}
\label{BandStructure}
\end{figure*}

For bulk calculations, a plane-wave cutoff of 60 Ry is used and the primitive unit cell is sampled with a Monkhorst k-mesh of $9 \times 9 \times 9$. Spin-orbit coupling is included in the calculations. The calculated band structure resulting from a bulk unit cell is shown in Fig. \ref{BandStructure}.

In order to calculate the surface spectrum and spin-orbit texture, a Wannier tight-binding model is generated from the $p_{x,y,z}$ orbitals of the O atoms and the $d_{xy,yz,zx}$ orbitals of Ta using the Wannier90 software package \cite{pizzi_wannier90_2020}. Following Ref. \cite{bruno_band_2019}, the 2DEG at the surface can be modeled through the introduction of a potential well at the surface in order to avoid the explicit introduction of symmetry breaking terms. The magnitude of this potential well is the only parameter tuned in the calculations and is fitted such that the results are in alignment with those of Ref. \cite{bruno_band_2019}. The surface spectra and spin-orbit texture is then calculated using the WannierTools software package \cite{wu_wanniertools_2018}. The results of the surface spectra calculations for the three crystal terminations are shown in Fig. \ref{fig:ktosurfacestates}. Our calculations indicated a maximum Rashba coefficient, $\alpha_{R} \approx$ 2 meV\AA, for the (111) and (110) surfaces, in line with what is found by Bruno \textit{et. al} \cite{bruno_band_2019} and $\alpha_{R} \approx$ 1 meV\AA \,for the (001) surface. Similarly, calculations of the spin texture reveal an out of plane contribution for the (111) surface which is absent in the (001) and (110) samples, as shown in Fig. \ref{fig:ktospintexture}. These results also agree with those published by Bruno \textit{et. al }\cite{bruno_band_2019}. 

\paragraph*{Presence of Finite Surface Berry-Curvature Density}
The results shown in Fig. \ref{fig:berrycurvature} of the main text were calculated for the (111) surface utilizing the same Wannier tight binding model with a surface potential well generated to study the 2DEG spectra. The Berry curvature is then calculated discretizing the Brillouin zone into a 200×200 grid of plaquettes. The wavefunction of all occupied states is then parallel transported around the plaquette following the procedure put forth by Fukui et. al \cite{fukui_chern_2005}.

We also calculated Berry curvature for the (001) and (110) surfaces. Neither demonstrated finite Berry curvature density in alignment with our expectation based on the symmetry of the system and lack of an out of plane spin texture.

\begin{acknowledgements}
This work was supported by the U.S. Department of Energy, Basic Energy Sciences, under Award No. DE-FG02-06ER46346. This work also made use of the NUFAB facility of Northwestern University’s NUANCE Center, which has received support from the SHyNE Resource (NSF ECCS-2025633), the IIN, and Northwestern’s MRSEC program (NSF DMR-1720139). Additional equipment support was provided by DURIP grant W911NF-20-1-0066.
\end{acknowledgements}

\bibliography{Nonlinear_Hall_Effect}
\end{document}